\documentstyle{article}\begin{document}

\title{Inflaton and dark matter in a random environment }
\author{Z. Haba \\
Institute of Theoretical Physics, University of Wroclaw,\\ 50-204
Wroclaw, Plac Maxa Borna 9, Poland\\
email:zhab@ift.uni.wroc.pl} \maketitle

\begin{abstract}
We consider a Lagrangian of interacting classical fields. We
divide the Lagrangian  into two parts. The first part is to
describe either the dark matter (DM) or the inflaton (IN)
depending on the choice of the self-interaction. The second part
constitutes  an environment of an infinite number of scalar fields
interacting linearly with the first part. We approximate the
environment by a white noise obtaining a Langevin equation.  We
show that  the resulting Fokker-Planck equation has  solutions
determining a relation between the diffusion constant, the
cosmological constant and the temperature. As a consequence of the
Langevin approximation the energy-momentum tensor of the dark
matter and the inflaton is not conserved. The compensating
energy-momentum tensor is interpreted as the dark energy (DE). We
insert the total energy-momentum in Einstein equations.  We show
that under special initial conditions Einstein equations have a
solution with a constant ratio of DM/DE and IN/DE
  densities.
\end{abstract}
\section{Introduction}
The $\Lambda$CDM model became the standard cosmological model
since the discovery of the universe acceleration
\cite{a1}\cite{a2}. It describes very well the large scale
structure of the universe. The formation of an early universe is
well explained by the inflationary models involving scalar fields
\cite{planck15}\cite{st}. However, a model describing universe
evolution from its hot early stage till the present day is still
missing. Such a model should reveal the nature of the mysterious
dark matter (DM), dark energy (DE) and the surprising relation
between various parameters in the $\Lambda$CDM model (the
coincidence problem, the cosmological constant problem). The role
of the inflaton field in the high energy physics is also obscure
\cite{shaposh}. Some  hints concerning the precise theory
connecting the different ingredients of the $\Lambda$CDM model may
come from the observations on the galactic scale, where the
$\Lambda$CDM model encounters some difficulties in explaining the
dark matter halos. In ref.\cite{stein}, where the difficulties are
reviewed, it has been suggested that DM self-interaction can solve
the problems. However, DM self-interaction itself leads to
difficulties, e.g., problems with an explanation of spiral halos
of DM \cite{spiral}. In our earlier papers
 \cite{cqg}\cite{hss}\cite{habarel} we introduced a new ingredient in
 cosmological models: a diffusive interaction between dark matter and dark energy.
 Our point of view  is that the dark energy
consists of some unknown particles and fields. They interact in an
unknown way with particles of the dark matter and the inflaton.
The result
 of the
interaction could be seen in a diffusive behaviour of the dark
matter. The diffusion effect does not depend on the details of the
interaction but only on its strength and "short memory" (Markovian
approximation). In this paper we follow an approach appearing in
many papers (see
\cite{peebdm}\cite{peebin}\cite{bolotin}\cite{bento}\cite{faraon}
and references quoted there) describing the dark matter and the
fields responsible for an inflation (inflatons) by scalar fields.
What is new in our model is an introduction of an environment of
an infinite set of scalar fields interacting with DM and the
inflaton (IN). In a limit of an infinite number of fields the dark
energy is described by a random ideal fluid. The model is built in
close analogy to the the well-known infinite oscillator model
\cite{ford}\cite{fordkac}\cite{kleinert} of Brownian motion. There
is some similarity of our model to the warm inflation \cite{warm}
in the use of the white noise in the description of
post-inflationary evolution. The aim of the model is to describe
the hot inflationary phase   as well as the late evolution with
the same set of  scalar fields. However, we do not discuss here
consequences of the model for inflation. We concentrate in this
paper on the late time dynamics. The early inflationary phase
could be treated within the quantized version of the model in the
Starobinsky-Vilenkin stochastic approximation
\cite{vilenkin}\cite{starob}. In this preliminary study we show
that the  time evolution described by the Langevin equation can
predict the relation between the diffusion constant, the
temperature and the Hubble constant (a version of the
fluctuation-dissipation theorem) as well as the relation between
DM density,inflaton density and  DE density. The plan of the paper
is as  follows.  In sec.2 we define the model in its heuristic
form and indicate its relation to the well-known
Starobinsky-Vilenkin model \cite{vilenkin}-\cite{starob}. In sec.3
we introduce a Langevin equation as a limit of  an
 infinite number of the environmental scalar fields. As a consequence of the Fokker-Planck equation resulting from
  the Langevin equation  we derive a relation
between the temperature, the diffusion constant and the Hubble
constant. In sec.4 we discuss the conservation law of the
energy-momentum. In order to preserve the conservation of the
total energy-momentum we have to introduce a compensating
energy-momentum interpreted as the dark energy. In sec.5 we suggest that
an approximate equation of state needed to close Friedmann
equations can be obtained by averaging either in time or over
configurations. The energy-momentum is inserted in the Einstein
equations in sec.6. We obtain a particular solution of these
equations which gives a constant ratio of DM/DE density and IN/DE
energy density. The solution is the fixed point of a dynamical
system discussed in \cite{hss}\cite{sz}.

\section{Scalar fields  interacting with an environment}
The CMB observations show that the universe was once in an
equilibrium state. The Hamiltonian dynamics of  scalar fields
usually discussed in the model of inflation do not equilibrate. We
can achieve an equilibration if the scalar field interacts with an
environment. We suggest a field theoretic model which is an
extension of the well-known oscillator model discussed in
\cite{ford}\cite{fordkac}\cite{kleinert}. We consider the
Lagrangian
\begin{equation}\begin{array}{l}
{\cal L}=\frac{1}{2}\partial_{\mu}\phi\partial^{\mu}\phi
-V(\phi)+\sum_{a}(\frac{1}{2}\partial_{\mu}\chi^{a}\partial^{\mu}\chi^{a}
-\frac{1}{2}m_{a}^{2}\chi^{a}\chi^{a}-\lambda_{a}\phi\chi^{a}).
\end{array}\end{equation}
Equations of motion read
\begin{equation}
g^{-\frac{1}{2}}\partial_{\mu}(g^{\frac{1}{2}}\partial^{\mu})\phi=-V^{\prime}-\sum_{a}\lambda_{a}\chi^{a},
\end{equation}
\begin{equation}
g^{-\frac{1}{2}}\partial_{\mu}(g^{\frac{1}{2}}\partial^{\mu})\chi^{a}+m_{a}^{2}\chi^{a}=-\lambda_{a}\phi,
\end{equation}where $g_{\mu\nu}$ is the metric tensor and $g=\vert \det[g_{\mu\nu}]\vert$.
 Inserting
the solution of eq.(3) in eq.(2) we obtain an equation of the form
\begin{equation}\begin{array}{l}
g^{-\frac{1}{2}}\partial_{\mu}(g^{\frac{1}{2}}\partial^{\mu})\phi+m^{2}\phi+V^{\prime}(\phi)
=-\int_{0}^{t}{\cal
K}(t,t^{\prime})\phi(t^{\prime})dt^{\prime}+W(\chi(0),\partial_{t}\chi(0)),
\end{array}\end{equation}where the kernel ${\cal K}$
is an evolution  kernel for the linear equation (3) and the noise
$W$  depends linearly on the initial conditions
$(\chi(0),\partial_{t}\chi(0))$ for the second order differential
equation (3). We could quantize the scalar field equations
(2)-(3). Then, the initial values $W(\chi(0),\partial_{t}\chi(0))$
 in eq.(4) will be the quantum fields. We could assume that these
fields are in the thermal state $\exp(-\beta H_{\chi}) $, where
$H_{\chi}$ is the quantum Hamiltonian of the $\chi$ fields and
$\frac{1}{\beta}$ is the temperature of the heat bath.
Subsequently, we can take the classical limit of the quantum field
theory. The first term on the rhs of eq.(4) describes a friction
coming from the environment. The second term is  the "noise"  from
the environment. It will have a certain probability distribution
in a quantum theory if the initial conditions are quantum fields (
or random thermal fields in the classical limit). In the limit of
an infinite number of fields with properly chosen masses and
couplings we obtain the quantum analogue of the white noise ( in
complete analogy to the model of refs.
\cite{ford}\cite{fordkac}\cite{kleinert} ). If we choose the
probability distribution as the classical limit of the quantum
thermal state $\exp(-\beta H_{\chi}) $, then we can obtain the
classical noise (discussed in the next section).

We consider in general the metric
\begin{equation}
ds^{2}=g_{\mu\nu}dx^{\mu}dx^{\nu}=g_{00}dt^{2}-g^{(3)}_{ij}dx^{i}dx^{j}.
\end{equation}Explicit formulas  will be derived for a flat FLWR expanding
metric
\begin{equation}
ds^{2}=dt^{2}-a^{2}d{\bf x}^{2},
\end{equation}
when $H=a^{-1}\partial_{t}a=$const (de Sitter space).

The well-known Starobinsky-Vilenkin stochastic equation
\cite{vilenkin} \cite{starob} does not come from an interaction
with an environment but could be obtained from a quantum version
of eq.(2) with $\lambda_{a}=0$. In such a case the field $\phi$ is
cut at high momentum exceeding the de Sitter horizon. We could
apply the Starobinsky-Vilenkin approximation by considering two
independent noises on the rhs of eq.(4): one describing the
quantum fluctuations of the $\phi$ field and another one coming
from an interaction with the $\chi^{a}$ fields. The
Starobinsky-Vilenkin approximation corresponds to an infinite
"friction" $H$ (or "slow rolling") . Then, neglecting the
$\partial_{t}^{2}\phi$ term in eq.(2) we obtain
\begin{equation}
3H\partial_{t}\phi=-V^{\prime}+W_{SV},
\end{equation}
where the Starobinsky-Vilenkin noise $W_{SV}$ has the correlations
determined by the quantum field (in the de Sitter space) cut at
high momenta
\begin{equation}
\langle W_{SV}(t,{\bf x})W_{SV}(t^{\prime},{\bf
x})\rangle=\frac{H^{3}}{4\pi^{2}}\delta(t-t^{\prime}).\end{equation}
The Starobinsky-Vilenkin equation is supposed to describe
fluctuations during inflation. We suggest that eqs.(2)-(3)
 with a proper choice of the potential $V$ can  describe the  time evolution
 of the inflaton embedded in
the environment of dark energy. The potential $V$ should agree
with the observational data
\cite{planck15}\cite{pot}\cite{odintsov}\cite{stein}. Then,
$V(\phi)$ must  fall at large $\phi$ in order not contradict the
observed outcome of the primordial nucleosynthesis \cite{peebin}.
Eq.(4) could also describe the dark matter with another choice of
$V(\phi)$. In particular, Higgs-type potentials are taken into
account \cite{peebdm}\cite{bertolami}. Models with a single field
and a single potential describing dark matter and inflation are
also discussed \cite{liddle}.

\section{A relation between the diffusion constant, temperature and the Hubble
constant} We restrict ourselves to classical field theory and
neglect the Starobinsky-Vilenkin noise. We need a formulation of
Eq.(4) with an infinite number of fields $\chi$ in a form which is
covariant under the change of coordinates on the manifold. This
leads  to a direct generalization of the Kramers equation
\cite{risken} (usually expressed in the phase space)
\begin{equation}
g^{-\frac{1}{2}}\partial_{\mu}(g^{\mu\nu}g^{\frac{1}{2}}\partial_{\nu})\phi+V^{\prime}(\phi)=\gamma
W.
\end{equation} Here, the noise
$W$ is Gaussian with mean zero and the correlation function
\begin{equation}
\langle
W(x)W(x^{\prime})\rangle=\delta_{g}(x,x^{\prime})=g^{-\frac{1}{2}}\delta(x,x^{\prime}),
\end{equation}where  the $\delta_{g}$-function on a manifold is defined by
\begin{displaymath}
\int dx\sqrt{g}\delta_{g}(x,x^{\prime})f(x)=f(x^{\prime})
\end{displaymath} and $\delta$ on the rhs of eq.(17) is understood as the one in local coordinates around
$x$. Eqs.(9)-(10) lead to the proper (intrinsic) formulation of
the stochastic wave equation on a manifold independent of the
choice of coordinates \cite{brzezniak}.
 For a flat expanding metric
we have
\begin{equation}
\partial_{t}^{2}\phi-a^{-2}\triangle\phi+3H\partial_{t}\phi+V^{\prime}(\phi)=\gamma W.
 \end{equation} In eq.(11) we have neglected the friction $\Gamma_{en}$
coming from the environment (present in eq.(4), in general
$3H\rightarrow 3H+\Gamma_{en}$). We rewrite eq.(11) as a system of
first order equations for $\phi$ and $\partial_{t}\phi$. The
Fokker-Planck equation for the stochastic system (11) is (we omit
the spatial derivatives in
eq.(11))\begin{equation}\begin{array}{l}
\partial_{t}P=\frac{\gamma^{2}}{2}\int d{\bf
x}g^{-\frac{1}{2}}\frac{\delta^{2}}{\delta\Pi({\bf
x})\delta\Pi({\bf x})}P\cr+3H\int d{\bf
x}\frac{\delta}{\delta\Pi({\bf x})}\Pi({\bf x})P-\int d{\bf
x}V^{\prime}(\phi({\bf x}))\frac{\delta}{\delta\Pi({\bf x})}P
+\int d{\bf x}\Pi({\bf x})\frac{\delta}{\delta\phi({\bf
x})}P.\end{array}\end{equation} We solve eq.(12) with the initial
condition
\begin{displaymath}
P(0)=\exp\Big(-\frac{\beta}{2}\int d{\bf
x}(g^{(3)}(0))^{\frac{1}{2}}\Pi^{2}\Big),
\end{displaymath}where $g^{(3)}=\det(g_{ij}^{(3)})$.
The  solution of eq.(12) for $V=0$ is
\begin{equation}
P(t)=L(t)\exp\Big(-\frac{1}{2}\alpha(t)\int d{\bf x}\Pi^{2}\Big)
\end{equation}
with
\begin{equation}
\alpha(t)=\sqrt{g}a^{3}\Big(\beta^{-1}\sqrt{g(0)}+\gamma^{2}\int_{0}^{t}a(s)^{3}ds\Big)^{-1}
\end{equation}and
\begin{displaymath}
L(t)=\exp\Big(-\int d{\bf x}\delta({\bf 0})\int_{0}^{t}
(\frac{\gamma^{2}}{2}(g(s))^{-\frac{1}{2}}\alpha(s)-3H(s))ds\Big)
\end{displaymath}
$L(t)$ contains infinities $\int d{\bf x}$ and $\delta({\bf 0})$
which would cancel  if  the rhs of eq.(12) was defined with a
proper point splitting. However, there is no need to tackle the
$L(t)$ factor because it drops out in the calculations of the
normalized expectation values with respect to the measure $P$.

 For a large time we have
\begin{displaymath}
\alpha(t)\simeq\sqrt{g}a^{3}\Big(\gamma^{2}\int_{0}^{t}a(s)^{3}ds\Big)^{-1}
\end{displaymath}
In de Sitter space $a(t)=\exp(Ht)$. Then, for a large time
\begin{displaymath}
\alpha(t)\rightarrow \sqrt{g}T_{\infty}^{-1}
\end{displaymath} where
\begin{equation}
T_{\infty}=\frac{\gamma^{2}}{3H},
\end{equation}
In the general metric (5) with $g_{ij}\simeq \exp(Ht)$ we would
get $T_{\infty}=T_{st}\sqrt{g_{00}}$ where $T_{st}$ is Tolman's
equilibrium temperature \cite{tolman} (see also \cite{landau}).

In classical field theory in a static background metric $P(0)$ is
the canonical Gibbs distribution determined by the maximum of the
entropy $S=-\int P\ln P$.  $P(t)$ has an interpretation as a Gibbs
probability distribution at time $t$.  We can define the
temperature of the state $P$ by the formula
\begin{equation}
\frac{1}{T}=\frac{\partial S}{\partial E} \end{equation} where the
energy  $E$(for $V=0$) is
\begin{equation}
E=\frac{1}{2}\langle \int d{\bf x}\sqrt{g^{(3)}}\Pi^{2}\rangle
\end{equation}
where the expectation value is with respect to $P$. It follows
from the formula (13) for $P$ and from eq.(16) that
\begin{equation}
\frac{1}{T}=\alpha(t)a^{-3}
\end{equation}

 We could determine the temperature in another way: as $\frac{1}{2}$ of the the mean value of the "kinetic
 energy" $\langle \int d{\bf x}\sqrt{g^{(3)}}
\frac{1}{2}\Pi(t,{\bf x})\Pi(t,{\bf x})\rangle$. For the
calculation of the temperature we can neglect the interaction and
the term $a^{-2}\triangle$ in eq.(11). Then, the solution of
eq.(11)  is
\begin{equation}\begin{array}{l}
\Pi(t)=\exp(-\int_{0}^{t}H(t^{\prime})dt^{\prime})\Pi(0)+\gamma\int_{0}^{t}\exp(-3\int_{s}^{t}H(t^{\prime})dt^{\prime})W(s)ds.
\end{array}\end{equation} We have
\begin{equation}\begin{array}{l}
\langle \Pi(x)\Pi(x^{\prime})\rangle=a(t)^{-6}a(0)^{6}\Pi(0,{\bf
x})\Pi(0,{\bf x}^{\prime})\cr+ \delta({\bf x}-{\bf
x}^{\prime})\gamma^{2}a(t)^{-6}\int_{0}^{t}a(s)^{3}ds\simeq
\delta({\bf x}-{\bf
x}^{\prime})\frac{\gamma^{2}}{3H(t)}\end{array}
\end{equation}
at large time (for a slowly varying  $H$ ). This result coincides
with eq.(15). We have got here the same relation (15) between the
temperature, diffusion constant and the Hubble constant as for a
particle diffusion in \cite{habarel} (for an exact equality we
need $\gamma^{2}=3\kappa^{2}$, where $\kappa^{2}$ is the diffusion
constant for  particle's diffusion). If  we assume that the
classical fields in eqs.(2)-(3) result as the classical limit of
quantum fields , which  are defined in the de Sitter space and
have a well-defined temperature $T_{\infty}$, then we have in
addition the requirement $T_{\infty}=\frac{H}{2\pi}$ (this
relation has been derived in \cite{figari}\cite{gibbons} as a
consequence of the periodicity in the imaginary time) . Hence, in
the de Sitter stage of the evolution
\begin{equation}
H^{2}=\frac{2\pi\gamma^{2}}{3}.
\end{equation}
The relation (18) for a general evolution $a(t)$ coincides with
the one discussed in \cite{cqg}\cite{hss} if DM equation of state
is $\tilde{w}=1$. The reason  for its applicability to this case
($ V=0$) will be explained in sec.6. Eqs. (15) and (21) can apply
separately at different stages of the universe evolution.
\section{The energy-momentum tensor of the interaction with an environment}
The total energy-momentum tensor resulting from the Lagrangian (1)
is conserved and could be inserted on the rhs of Einstein
equations. However, if we replace the infinite set of fields
$\chi^{a}$ by the noise (as in eq.(9)) then the conservation law
for the energy-momentum tensor $T^{\mu\nu}$ of the field $\phi$
fails. We have to compensate in the energy-momentum the
replacement of the fields $\chi^{a}$ by the noise by means of a
compensating energy-momentum $T_{\Lambda}$ which we interprete as
dark energy.
 Now, the conserved energy-momentum tensor
$T_{tot}^{\mu\nu}$ is
\begin{equation} T_{tot}^{\mu\nu}=T^{\mu\nu}+T^{\mu\nu}_{\Lambda}.\end{equation}
From the conservation law
\begin{equation}
(T_{\Lambda}^{\mu\nu})_{;\mu}=-(T^{\mu\nu})_{;\mu}.\end{equation}
 Many models of  dark energy can be described by the energy-momentum
of an ideal fluid
\begin{equation}
T^{\mu\nu}_{\Lambda}=(\rho_{\Lambda}+p_{\Lambda})u^{\mu}u^{\nu}-g^{\mu\nu}p_{\Lambda},
\end{equation}where $\rho$ is the energy density and $p$ is the
pressure.  The velocity $u^{\mu}$ satisfies the normalization
condition
\begin{displaymath}  g_{\mu\nu}u^{\mu}u^{\nu}=1.\end{displaymath}

For the scalar field we have the representation (24) with
\begin{equation}
u^{\mu}=\partial^{\mu}\phi(\partial^{\sigma}\phi\partial_{\sigma}\phi)^{-\frac{1}{2}},
\end{equation}
\begin{equation}
\rho+p=\partial^{\sigma}\phi\partial_{\sigma}\phi,
\end{equation}
\begin{equation}p=\frac{1}{2}\partial^{\sigma}\phi\partial_{\sigma}\phi-V.
\end{equation}
We have from eq.(9) (with $T^{\mu\nu}$ defined by eqs.(24)-(25))
\begin{equation}
(T^{\mu\nu})_{;\mu}=\gamma\partial^{\nu}\phi W
\end{equation}

 The divergence equation (28) in a homogeneous
metric  in the frame $u=(1,{\bf 0})$ (spatial homogeneity of
$\phi$) gives
\begin{equation}
\partial_{t}\rho+3(1+\tilde{w})H\rho=\gamma W\partial_{t}\phi
,\end{equation}where \begin{equation} \tilde{w}=
(\frac{1}{2}\Pi^{2}-V)(\frac{1}{2}\Pi^{2}+V)^{-1}. \end{equation}
Then, for the compensating energy density we have (from eq.(23))
\begin{equation}
\partial_{t}\rho_{\Lambda}+3H(1+w)\rho_{\Lambda}=-\gamma W\partial_{t}\phi,
\end{equation}where $H=a^{-1}\partial_{t}a$ and
\begin{displaymath}
w=\frac{p_{\Lambda}}{\rho_{\Lambda}}.\end{displaymath} The
solution of eq.(31) with a constant $w$ is
\begin{equation}\begin{array}{l} \rho_{\Lambda}(t)=a^{-3(1+w)}\sigma_{\Lambda}-\gamma
a(t)^{-3(1+w)}\int_{t_{0}}^{t}a(s)^{3+3w}W(s)\partial_{s}\phi(s)ds.\end{array}\end{equation}
$\sigma_{\Lambda} $ is a constant such that
\begin{displaymath}
\rho_{\Lambda}(t_{0})=a(t_{0})^{-3(1+w)}\sigma_{\Lambda}.
\end{displaymath}
Various models with  a non-zero term on the rhs of eq.(29)
(interpreted as a time derivative of the cosmological term) have
been discussed
\cite{review}\cite{sola}\cite{bas}\cite{ss}\cite{nano}. In the
model (29) the cosmological term  is random.
\section{Averaging over fields}
For scalar fields the equation of state $p=\tilde{w}(\phi)\rho$
depends  on the field $\phi$. In a Hamiltonian system $\phi$
varies frequently in time   without achieving any equilibrium.
After adding the noise we may expect a smooth behaviour and an
equilibration.  Let us calculate (using equations of motion (11)
with an omission of spatial derivatives)
\begin{equation}\begin{array}{l}
\partial_{t}^{2}(a^{\frac{3}{2}}\phi^{2})=(\frac{3}{2}\partial_{t}H+\frac{9}{4}H^{2})a^{\frac{3}{2}}\phi^{2}
+2a^{\frac{3}{2}}((\partial_{t}\phi)^{2}-\phi
V^{\prime})+\gamma\langle \phi W\rangle.\end{array}\end{equation}
Taking the time average we obtain zero on the lhs. The term
$\langle \phi W \rangle=0$ for a causal
("non-anticipating"\cite{nonanti}) solution of eq.(11). Then, the
rhs gives the relation between kinetic energy and the potential
energy, i.e., the virial theorem.

 The environment can   make the system ergodic. In an ergodic system the time average is equal to the ensemble
average. We can express eq.(33) as
\begin{equation}\begin{array}{l}
\langle(\frac{3}{2}\partial_{t}H+\frac{9}{4}H^{2})\phi^{2}+
2(\partial_{t}\phi)^{2})\rangle=\langle\phi
V^{\prime}\rangle,\end{array}\end{equation}
where the average is understood either in time or over configurations.
 We suggest that we can make the
approximation for late time evolutions
\begin{equation}
\tilde{w}\simeq\tilde{w}_{av}=\langle p\rangle \langle
\rho\rangle^{-1}.
\end{equation}
Let us calculate $\tilde{w}_{av}$ in some special cases using for
the average over configurations the measure (12) defined by $ P$,
corresponding to $H=0$, then
\begin{displaymath}
P=\exp\Big(-\int d{\bf x}(\frac{1}{2}\Pi^{2}+V(\phi))\Big).
\end{displaymath}
 If $V(\phi)=\mu^{2}\phi^{2}$ then
\begin{equation} \tilde{w}_{av}=0.\end{equation} The result (36)
agrees with the usual assumption $\tilde{w}=0$ for massive
bodies.We would get the same result from the virial theorem (34)
(with $H=0$). For $V(\phi)=\mu\phi^{4}$ we obtain
 \begin{equation}
\tilde{w}_{av}=\frac{1}{3}\end{equation} from the average  over
configurations as well as from the virial theorem (with $H=0$)(34)
. The value $\frac{1}{3}$ is equivalent to the condition
$T^{\mu}_{\mu}=0$ resulting from conformal invariance of field
theory. The $\phi^{4}$ theory could be considered as a version of
the Brans-Dicke model (with the cosmological term) which is
conformal invariant.

In general, for $V=\mu\vert\phi\vert^{k}$ by a calculation of
elementary integrals we obtain from the average with respect to
the measure defined by $ P$
\begin{equation}\tilde{w}_{av}=\frac{k-2}{k+2}
\end{equation} Hence, $\tilde{w}\rightarrow 1$
for $k\rightarrow \infty$ and $\tilde{w}\rightarrow -1$ for small
$k $ (eq.(38) breaks down for $ k=0$, as if $V=0$ then obviously
$\tilde{w}=0$ from eq.(30)). For $k\geq 2$ we get from eq.(38)
$0\leq w_{av}<1$ as a possible value of $\tilde{w}_{av}$ for the
dark matter in the approach of \cite{peebdm}. For the inflaton
Peebles and Ratra suggest the behaviour $V(\phi)\simeq
\vert\phi\vert^{-\alpha}$ with $\alpha >0$. Hence,
$\tilde{w}_{av}=\frac{\alpha+2}{\alpha-2}$ which gives
$\tilde{w}_{av}>1$ if $\alpha >2$ and $\tilde{w}_{av}<0$ if
$0<\alpha <2$.

From the virial theorem (34) we obtain
\begin{equation}
\tilde{w}_{vir}=\Big(k-2-(\frac{3}{4}\partial_{t}H+\frac{9}{8}H^{2})k\Big)
\Big(k+2-(\frac{3}{4}\partial_{t}H+\frac{9}{8}H^{2})k\Big)^{-1}.
\end{equation}

The H-correction does not change our conclusion concerning the
limits of large and small $k$. Note that if $a\simeq t^{\sigma}$
then
$(\frac{3}{4}\partial_{t}H+\frac{9}{8}H^{2})=\frac{3}{2t^{2}}(\sigma-\frac{3}{2})\sigma$.
Hence, it is negative for non-accelerating cosmologies and
decreasing in time for a power law expansion.

\section{Einstein equations}
Einstein equations are written in the form
\begin{equation}
 R^{\mu\nu}-\frac{1}{2}g^{\mu\nu}R=8\pi G
T_{tot}^{\mu\nu}.
\end{equation} where $G$ is the Newton constant.
The energy-momentum tensor of sec.4 is a random variable. Hence,
eq.(40) describes a random evolution. We are not prepared at the
moment to discuss the general solutions of eq.(40). We take the
mean value on the rhs of eq.(40) treating  $a(t)$ as deterministic
( this is a conventional approach to the classical approximation
of the quantum energy-momentum). In principle, we could solve the
wave equation (9) and determine $T$ and $T_{\Lambda}$ as functions
of $a(t)$ from eqs.(29)-(31) (we need to know $\tilde{w}$ either
from the exact solution of the wave equation or by averaging of
sec.5) . Then, calculation of the expectation value of $T_{tot}$
over the noise $W$ would determine the rhs of Einstein equations
(40). It is difficult to do it for non-linear equations. We
restrict ourselves here to the simplified case  when $V=0$ and the
spatial derivatives in eq.(11) are neglected. Then, $ \rho=
T^{00}=\frac{1}{2}\Pi^{2}$. This energy density and its
expectation value $\rho_{av}$ has been calculated in eq.(20).
$\rho_{av}=\langle T^{00}\rangle$ solves the equation
\begin{equation}
\partial_{t}\rho_{av}+3H(1+\tilde{w})\rho_{av}=\gamma^{2}a^{-3}
\end{equation}with $ \tilde{w}=1$ as expected from eq.(30) with $V=0$.
Then, the compensating energy density $\rho_{\Lambda}^{av}=\langle
T^{00}_{\Lambda}\rangle$ (on the basis of eq.(23)) satisfies the
equation
\begin{equation}
\partial_{t}\rho_{\Lambda}^{av}+3H(1+w)\rho_{\Lambda}^{av}=-\gamma^{2}a^{-3}.\end{equation}
The Friedman equation in the FRLW flat metric (6) reads
\begin{equation}\begin{array}{l}
H^{2}=\frac{8\pi
G}{3}(\rho_{av}+\rho_{\Lambda}^{av}).\end{array}\end{equation}
Inserting the solutions of eqs.(41)-(42) in eq.(43) we obtain
\begin{equation}\begin{array}{l} H^{2}=\frac{8\pi
G}{3}\Big(a^{-3(1+\tilde{w})}\sigma +\gamma^{2}
a(t)^{-3(1+\tilde{w})}\int_{t_{0}}^{t}a(s)^{3\tilde{w}}ds+a^{-3(1+w)}\sigma_{\Lambda}
\cr-\gamma^{2}
a(t)^{-3(1+w)}\int_{t_{0}}^{t}a(s)^{3w}\Big).\end{array}\end{equation}
We have got the Friedmann equation which coincides with the one
for a diffusive matter in \cite{hss}\cite{habarel}. This
coincidence may be a consequence of the relation between the
energy-momentum of quantum fields and the energy-momentum of
particles which can be established on the basis of the Wigner
function formalism. Although in our simplified model with $V=0$ we
have $\tilde{w}=1$ (from eq.(30)) we believe that eq.(44) can be
also applied to $V\neq 0$ by an insertion of a constant
$\tilde{w}$ by means of the method of averaging of sec.5. We have
studied eq.(44) numerically  in \cite{hss} (see also \cite{sz}) in
order to explore the relation between $\rho$ and $\rho_{\Lambda} $
(the coincidence problem). For this purpose we formulated
eqs.(41)-(43) as a closed dynamical system. It comes out that the
fixed point of the dynamical system of refs.\cite{hss}\cite{sz}
coincides with the linear evolution of $a$ \begin{equation}
a(t)=a_{0}+\lambda t
\end{equation}(a comparison of the general numerical solution of eqs.(41)-(43) with observations is studied in
\cite{hss}\cite{sz}; a consistency of a linear evolution with
observations is discussed in \cite{n1}\cite{n2}).

In fact, we can show directly that (45) is the solution of eq.(44)
for some special initial conditions
$(a_{0},\sigma,\sigma_{\Lambda}$). If we set $t= t_{0}$ in eq.(44)
then we obtain
\begin{equation}\begin{array}{l}
\lambda^{2}=\frac{8\pi
G}{3}\Big(a_{0}^{-3(1+\tilde{w})}\sigma+a_{0}^{-3(1+w)}\sigma_{\Lambda}\Big).
\end{array}\end{equation}
After an integration over $s$ in eq.(44) (we set $t_{0}=0$)  with
the use of eq.(45) eq.(44) reads
\begin{equation}\begin{array}{l}
\frac{3}{8\pi G}\lambda^{2}a(t)^{-2}=\sigma
 a^{-3(1+\tilde{w})}+
\frac{\gamma^{2}}{\lambda(1+3\tilde{w})}(a(t)^{-2}-a_{0}^{1+3\tilde{w}}a(t)^{-3(1+\tilde{w})})\cr+\sigma_{\Lambda}
a^{-3(1+w)}
-\frac{\gamma^{2}}{\lambda(3w+1)}a(t)^{-2}+\frac{\gamma}{\lambda(3w+1)}a_{0}^{3w+1}a(t)^{-3(1+w)}.
\end{array}\end{equation}
If we wish a solution for arbitrary $w$ and $\tilde{w}$ then the
terms $a^{-3(1+\tilde{w})}$ and $a^{-3(1+w)}$ in eq.(47) should
cancel. Hence,
\begin{equation}
\sigma_{\Lambda}=-\frac{\gamma^{2}}{\lambda}(1+3w)^{-1}a_{0}^{1+3w}
\end{equation}and
\begin{equation}
\sigma=\frac{\gamma^{2}}{\lambda}(1+3\tilde{w})^{-1}a_{0}^{1+3\tilde{w}}.
\end{equation}From eq.(48) it follows that $1+3w<0$ if $\sigma_{\Lambda}$ is to be non-negative.
If the terms at $a^{-2}$ in eq.(47) are to cancel then
\begin{equation} \frac{1}{8\pi G}\lambda^{3}
=\gamma^{2}(w-\tilde{w})(1+3w)^{-1}(1+3\tilde{w})^{-1}.
\end{equation}
Together with eqs.(48)-(49) eq.(50) is the same as eq.(46).
Eq.(50) determines $\lambda$, then solving eqs.(46),(48)-(49) we
determine the initial conditions $\sigma$,
 $\sigma_{\Lambda}$ and $a_{0}$. We obtain $a_{0}=1$ and
 $\sigma=\rho(0)=-(1+3w)(1+3\tilde{w})^{-1}\rho_{\Lambda}(0)\simeq (8\pi G)^{-\frac{1}{3}}\gamma^{\frac{2}{3}}$.
  Using the initial conditions (48)-(49) we
can calculate the dark energy from eq.(42)
\begin{equation}
\rho_{\Lambda}(t)=-\frac{\gamma^{2}}{\lambda}(1+3w)^{-1}a(\tau)^{-2}
\end{equation}
In the units $8\pi G =1$ with $\tilde{w}=0$ and $w=-1$ we obtain
from eq.(51)  the cosmological term at large time
$\Lambda(t)\simeq 0.25 t^{-2}$ (this behaviour is tested against
observations in \cite{ss}). The density of the dark matter follows
from eq.(41)
\begin{equation}
\rho(t)=\frac{\gamma^{2}}{\lambda}(1+3\tilde{w})^{-1}a(\tau)^{-2}.
\end{equation}
Hence, \begin{equation}
\rho(t)\rho_{\Lambda}(t)^{-1}=-(1+3\tilde{w})^{-1}(1+3w).
\end{equation}For the
relativistic models of \cite{hss}\cite{sz}, when the dark matter
density is determined by a phase space distribution we have $
0<\tilde{w}\leq \frac{1}{3}$. Hence,
$\frac{1}{2}<\rho_{DM}(t)\rho_{\Lambda}(t)^{-1}\leq 1$ for $w=-1$.
We have a greater range of $\rho(t)\rho_{\Lambda}(t)^{-1}$ for
scalar field theories. So, for the inflaton of \cite{peebin} with
$k=-\alpha$ in eq.(38) we can have arbitrarily large $\tilde{w}$
hence $\rho_{IN}\rho_{\Lambda}^{-1}$ can be arbitrarily small
whereas for the dark matter with $0\leq \tilde{w}<1$ (when $k\geq
2$) $\rho_{DM}\rho_{\Lambda}^{-1}$ varies in the interval
$(\frac{1}{8},\frac{1}{2}]$(for $w=-1$). The general, solution of
eqs.(41)-(43) gives the $\rho\rho_{\Lambda}^{-1}$ ratio which
varies in time \cite{hss}\cite{sz}. We can show that if the
conservation equations (23) are satisfied and we require that
$\rho(t)\rho_{\Lambda}(t)^{-1}=$const then $a(t)$ satisfying
Friedmann  equations must be a linear function of $t$.

\section{Discussion and summary}
We have drawn some conclusions resulting from the assumption that
the relativistic  IN/DE or DM/DE interaction has a diffusive
character.  We have obtained earlier  related results in a model
of a particle DM/DE interaction resulting from a relativistic
diffusion. It seems that some aspects of the diffusive
interactions are independent of peculiarities of the models. The
diffusive dynamics leads to a fixed ratio  between IN/DE and DE/DM
densities. The relation between the diffusion constant and the
temperature is a version of fluctuation-dissipation theorem
well-known in statistical physics, whereas the relation between
temperature and the cosmological constant is connected with the
definition of temperature in quantum field theory. These relations
can supply a new look at the cosmological constant and the
coincidence problems. The diffusion constant could  be measurable
by an observation of the dark matter dynamics through its effect
on the luminous matter. The inflaton density can be related to
dark energy and some parameters of the CMB spectrum. So, in
principle, the formulae (15) and (21) are verifiable. The model is
supposed to be applicable in a large range of time. In the same
model we could apply the Starobinsky-Vilenkin approximation
describing the inflaton fluctuations in a unified way for the
inflation era as well as at large time.

The connection between the Hubble constant and the diffusion
constant suggests another explanation of the origin of the
cosmological term \cite{wein}. Its present small value comes from
the  energy loss of the environment of DE (energy gain of DM). We
have compared the diffusive dynamics (resulting from $\gamma\neq
0$) with observations in \cite{hss}. The relations (15) and (21)
could be tested in observations if we could estimate the diffusion
constant $\gamma^{2}$ and dark matter temperature on the basis of
the motion of the luminous matter. From eq.(18) it follows that
the temperature of the dark matter may increase during the
expansion. Such a conclusion has also been derived from
thermodynamics of the dark energy in \cite{lima}\cite{pavon}.We
have good estimates of the DM density on the basis of rotational
curves \cite{curves} and lensing observations \cite{lensing}. We
would need either a time dependence of the DM density or an
independent way to determine DM velocities. A dispersion of the
velocities could be applied to measure the temperature of the dark
matter \cite{demianski}. Another source of information on the DM
phase space distribution could come from computer simulations of
the formation of DM halos \cite{halosim}. It is known that
non-interacting DM is unable to describe the galactic DM halos
\cite{stein}. A dissipative component of
 the dark matter may be necessary \cite{dissip}. Some estimates on
 the diffusion constant might come from the heat transfer inside the
 halo \cite{heat} if  the heat diffusion was a result of DM diffusion in an environment of  DE.
In conclusion, the field theoretic model of the diffusive DM/DE
interaction suggested by the relativistic field theory of particle
physics can lead to testable consequences on the basis of numerous
observational data.

 {\bf Acknowledgements} The paper has been completed during my stay in Centro de Ciencias de
 Pablo Pascual in Benasque and was reported during the conference Cosmology and the Quantum Vacuum
 held there.
 The author is grateful for a pleasant atmosphere at this meeting. Stimulating discussions  with Marek Szydlowski are gratefully
 acknowledged. The research is supported by the NCN
grant DEC-2013/09/BST2/03455.

\end{document}